\documentclass[prl,twocolumn,showpacs,amsmath,amssymb]{revtex4}
\usepackage{graphicx}

\begin{document}
\title{Electron-electron interactions in the conductivity of graphene}
\author{A.~A.~Kozikov$^{1}$}
\author{A.~K.~Savchenko$^{1}$}
\author{B.~N.~Narozhny$^{2}$}
\author{A.~V.~Shytov$^{1}$}
\affiliation{$^{1}$School of Physics,
University of Exeter, EX4 4QL, U.K.}
\affiliation{$^{2}$Institut f\"{u}r Theorie der Kondensierten Materie,
Karlsruher Institut f\"{u}r Technologie, 76128 Karlsruhe, Germany}

\begin{abstract}
The effect of electron-electron interaction on the low-temperature conductivity of graphene is investigated experimentally.
Unlike in other two-dimensional systems, the electron-electron interaction correction in graphene is sensitive to the details of disorder.
A new temperature regime of the interaction correction is observed where quantum interference is suppressed by intra-valley scattering. We determine the value of the interaction parameter, $F_0^{\sigma}\simeq-0.1$, and show that its small value is due to the chiral nature of interacting electrons.
\end{abstract}

\maketitle

The low-temperature behaviour of the resistance of electron systems is determined by quantum effects. Two distinct phenomena are responsible for this: quantum interference of electron waves scattered by impurities (weak localisation) WL \cite{Bergman},  and electron-electron interaction in the presence of disorder, EEI  \cite{AltshulerAronov}. The WL correction is used to study electron dephasing, while the EEI correction, which is not sensitive to dephasing,  has been used to probe the dynamics of interacting electrons, e.g. \cite{ZNA, Proskuryakov1, Shashkin, Gershenson1, Vitkalov, Gershenson}. In graphene, the charge carriers are chiral and located in two valleys. As a result, WL is sensitive not only to inelastic (phase breaking) scattering, but also to a number of elastic scattering mechanisms \cite{Ando,McCannPRL06}. For this reason, the WL correction to the conductivity can be either negative or positive, depending on the experimental conditions \cite{Tikhonenko, TikhonenkoWAL}.

So far the effects of EEI on the low-temperature conductivity of graphene have not been studied experimentally, and only the high-temperature ballistic regime was analysed theoretically \cite{Cheianov}. It was predicted that at $k_BT\tau_p > 1$, where $\tau_p$ is the momentum relaxation time, the EEI correction is determined by coherent backscattering on a single impurity, which in graphene is suppressed due to the chirality of charge carriers \cite{Ando98}. As a result, the EEI correction can only occur due to scattering on atomically sharp defects and is expected to have a universal form which is independent of the details of the electron-electron interaction. In addition, the ballistic regime in graphene can only be realised at relatively high temperatures where the EEI effect has to be separated from strong effects of electron-phonon scattering. Thus the ballistic regime of EEI is not expected to be promising for the study of electron interactions in graphene.

In the diffusive regime, $k_BT \tau_p<1$, interacting electrons in graphene
scatter on multiple impurities, so that backscattering is less important. Hence
the EEI correction will contain information about the details of interaction. In this work we study the EEI effect on the conductivity of graphene
in this regime. To separate the corrections due to EEI and
WL, we combine measurements of the temperature dependence of the conductivity with studies of
magnetoresistance (MR). We show that our results are described by a logarithmic correction to the conductivity
\cite{AltshulerAronov}:
\begin{eqnarray}
\delta\sigma^{\rm EEI}(T)=
-A(F_0^\sigma)\frac{e^{2}}{2\pi^{2}\hbar}\ln\frac{\hbar}{k_BT\tau_p}~.
\label{eqn:EEI}
\end{eqnarray}
Here the coefficient $A(F_0^\sigma)$ is determined by the strength of interaction and the symmetry of electron states. We show that in graphene there is a new regime of EEI and find the value of the interaction parameter $F_0^\sigma$.
\begin{figure}
\includegraphics[width=\columnwidth]{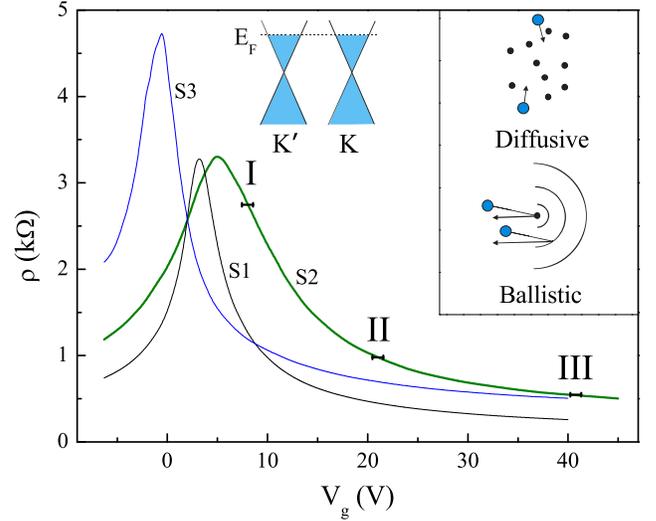}
\caption{ (Color online). The gate voltage dependence of the
resistivity for samples S1, S2 and S3. The bars
(shown for S2) indicate three studied regions. Insets: graphene
band structure with two valleys, and a schematic illustration of the regimes of interaction between two electrons that are scattered by impurities.}
\label{fig:one}
\end{figure}

Three samples with Hall-bar geometry (S1, S2 and S3) are fabricated by
mechanical exfoliation of graphite on Si/SiO$_{2}$ substrates
\cite{Monolayer}.  Sample parameters are shown in Table 1. Quantum Hall effect measurements were performed to
verify that the samples are monolayers \cite{Monolayer}. Figure \ref{fig:one} shows the resistivity $\rho$ as a function of the
gate voltage for the three samples. The bars indicate three regions of
the gate voltage where $\rho(T)$ was measured in all samples: $V_g=$ 3,
16 and 36 V with respect to the Dirac point.

\begin{figure}
\includegraphics[width=\columnwidth]{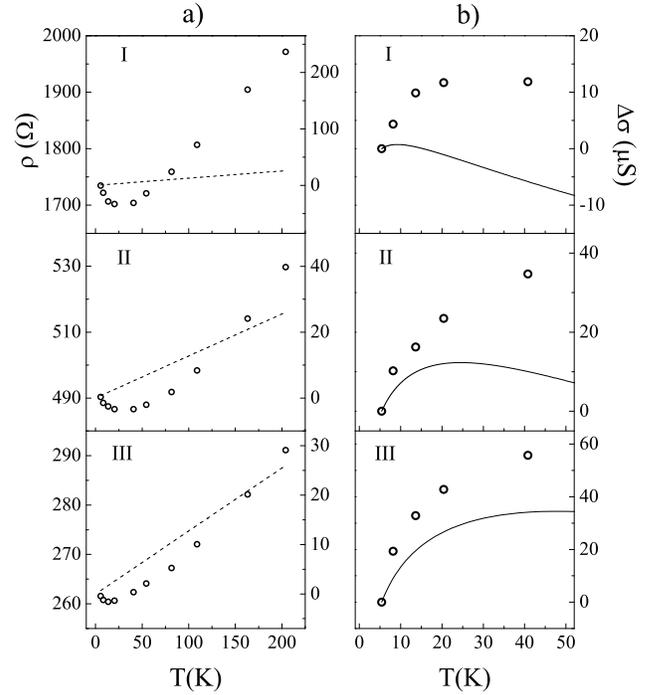}
\caption{ (a) The resistivity as a function of temperature for sample
S1 for three regions of carrier density. The dashed line is the acoustic phonon
contribution calculated using Eq.(\ref{eqn:phonons}) (the right-hand side axis).  (b) The conductivity after the phonon
contribution has been subtracted. Solid lines show the WL correction found from
Eq.(\ref{eqn:WL}). }
\label{fig:two}
\end{figure}

\begin{table}[t]
\begin{tabular}{c|ccc|c|c}
\multicolumn{1}{c}{} & \multicolumn{3}{c}{S1} & \multicolumn{1}{c}{S2} & \multicolumn{1}{c}{S3}\\
\hline
Region & $\mu$ & $\tau_i$& $\tau_*$ & $\mu$ & $\mu$ \\
\hline
I & 17500 &$~$14$~$& $~$0.45&$~$9300 &$~$12500  \\
II & 11500  &3&0.3& $~$5400 &$~$11000   \\
III &9700$~$&1 & $~$0.35& $~$4500 &9500  \\
\end{tabular}
\caption{Electron mobility $\mu$ (in cm$^2$V$^{-1}$s$^{-1}$ ) for three
regions of the carrier density in samples S1, S2 and S3. Characteristic scattering times $\tau_i$ and $\tau_*$ (in ps) are also shown for sample S1.}
\label{tab:Parameters}
\end{table}
Figure \ref{fig:two}a shows $\rho(T)$ in sample S1 in the temperature
range $T$= 5 - 200 K. The increase of the resistivity at high
temperatures can be partially ascribed to the effect of acoustic phonon scattering on the
classical conductivity \cite{Fuhrer, Morozov}, which is shown by the
dashed line \cite{Acoustic}:
\begin{eqnarray}
\rho_{ph}(T)=\left(\frac{h}{e^{2}}\right)
\frac{\pi^{2}D_{a}^{2}k_{B}T}{2h^{2}\rho_{s}\upsilon_{ph}^{2}\upsilon_{F}^{2}}~, \label{eqn:phonons}
\end{eqnarray}
where $D_a$=18 eV (as determined from the analysis of the classical conductivity \cite{Fuhrer}) is the deformation potential, $\rho_s=7.6\times10^{-7}~$kg~m$^{-2}$ is the
density of graphene, $\upsilon_{ph}=2\times10^4~$m~s$^{-1}$ is the speed of sound,
$\upsilon_{F}=10^6~$m~s$^{-1}$ is the Fermi velocity of carriers. We have subtracted the phonon contribution from the experimental dependence $\rho(T)$. The analysis has been limited to the range $T\leq$ 50 K in order to rule out other types of phonons at higher temperatures \cite{Fuhrer, Morozov}. The resulting
quantum correction to the conductivity is shown for all regions in Fig. \ref{fig:two}b as $\Delta \sigma(T)=\sigma (T)-\sigma (T_{0})$, where
$T_{0}$ is the lowest studied temperature, $\sigma (T)$ = [$\rho(T)$ -
$\rho_{ph}(T)$]$^{-1}$.

The separation of the EEI corrections from the WL contribution has been
performed by two methods. For samples S1 and S3 the low-field perpendicular
magnetoresistance has been measured, in order to determine the characteristic times responsible for WL: the inelastic dephasing time
$\tau_\varphi(T)$, the elastic time of inter-valley scattering $\tau_{i}$, and the elastic time $\tau_{*}$
which describes intra-valley suppression of quantum interference (due to
topological defects and `trigonal' warping of the energy spectrum
\cite{McCannPRL06}). (This analysis is done following the method described in
\cite{Tikhonenko, TikhonenkoWAL}.) These times are used to determine the WL
correction $\delta\sigma^{\rm WL}(T)$ \cite{McCannPRL06},
\begin{eqnarray}
\label{eqn:WL}
&&
\delta\sigma^{\rm WL}(T)=
-\frac{e^{2}}{2\pi^{2}\hbar}
\Bigg[\ln\left(1+2\tau_{\varphi}(T)/\tau_{i}\right)-
\\
&&
\quad\quad\quad\quad\quad
-2\ln\left(\frac{\tau_{\varphi}(T)/\tau_{p}}
{1+\tau_{\varphi}(T)/\tau_{i} + \tau_{\varphi}(T)/\tau_* }\Bigg)
\right]~,
\nonumber
\end{eqnarray}
which is then subtracted.
In sample S2, the EEI correction
has been isolated by suppressing WL by a perpendicular magnetic field which is still too small to affect the EEI
correction \cite{Bergman}. Both methods lead to close results for the magnitude of
the EEI correction in the studied samples.

The solid line in Fig. \ref{fig:two}b shows the WL correction to
the conductivity, $\Delta\delta\sigma^{\rm WL}(T)$ = $\delta\sigma^{\rm WL}(T)$ -
$\delta\sigma^{\rm WL}(T_0)$, found from the analysis of the magnetoresistance using the first method. One can see that the two types of quantum correction, WL and EEI, are of similar magnitude. The solid
lines show clearly that in regions I and II there is a transition from
weak localisation, an increase of $\Delta \sigma (T)$, to antilocalisation, a decrease of $\Delta \sigma (T)$. (Earlier, such a transition was detected in the change of the sign of MR \cite{TikhonenkoWAL}, with the transition temperatures of $\sim$ 10 K in region I and $\sim$ 25 K in
region II, which is in agreement with this experiment.)

\begin{figure}
\includegraphics[width=\columnwidth]{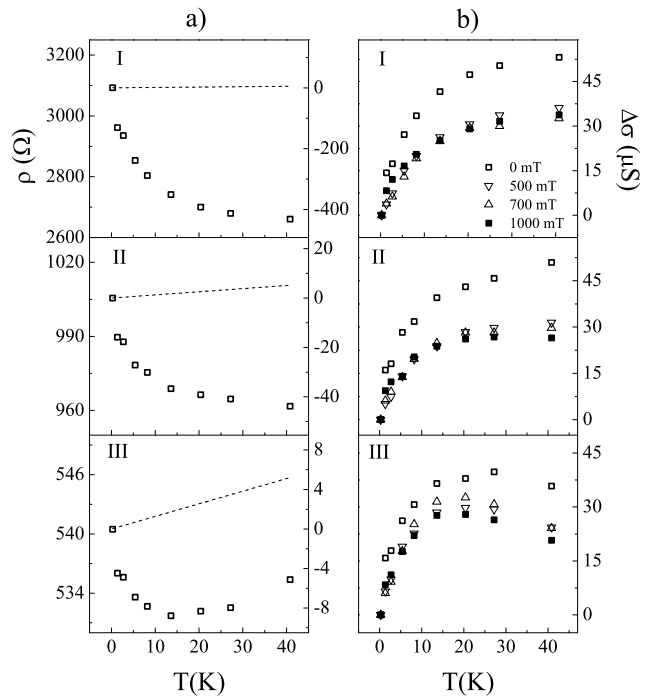}
\caption{ (a) The resistivity as a function of temperature for sample
S2, shown for three regions. The dashed line is the acoustic phonon
contribution calculated using Eq.(\ref{eqn:phonons}) (the right-hand side axis).
 (b) The conductivity $\Delta \sigma
(T)=\sigma (T)-\sigma (T_{0})$ at different magnetic fields (the contribution of
acoustic phonons has been subtracted).}
\label{fig:three}
\end{figure}

Figure \ref{fig:three}a shows the temperature dependence of the
resistivity of sample S2 in the
temperature range 0.25 - 40 K. First, the phonon contribution,
Eq.(\ref{eqn:phonons}), shown by the dashed line is
subtracted in regions II and III. The remaining quantum contribution to the conductivity is presented in Fig. \ref{fig:three}b, for different magnetic fields. One can see that with increasing $B$ there is a decrease in the slope of the temperature dependence until a saturation is reached.  This is a signature that the WL correction has been suppressed while the EEI correction is not affected by magnetic field.

Indeed, the suppression of WL is expected at fields which are much larger than
the so-called `transport' field $B_{tr}$ = $\hbar/2el_{p}^{2}$, where
$l_{p}$ is the mean free path \cite{Btr}. For sample S2 the
values of $B_{tr}$ are 120, 70 and 45 mT for regions I, II and III,
respectively, and therefore it is not surprising that WL appears
to be suppressed at $B=1$ T, Fig.~\ref{fig:three}b. On the other hand, the effect of the magnetic field on the EEI correction
is due to the Zeeman splitting of the triplet
`channel' and is expected at higher fields, $g^{*}\mu_{B}B>k_BT$
\cite{AltshulerAronov}, where $g^{*}$ is the
Land\'{e} g-factor, and $\mu_{B}$ is the Bohr magneton. For
the g-factor in graphene $\sim$2 \cite{Lande} and temperatures above 1 K, this condition is satisfied at fields higher than 1 T.

\begin{figure}[t]
\includegraphics[width=\columnwidth]{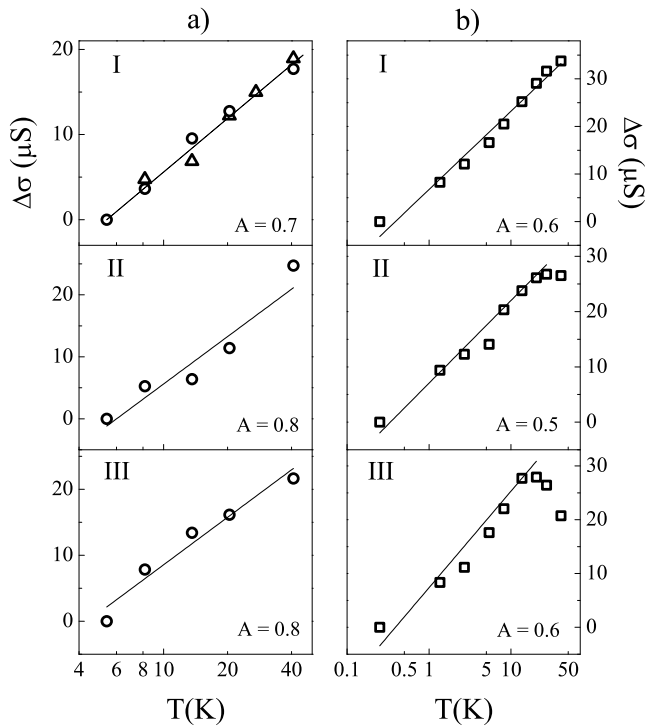}
\caption{ The electron-electron interaction correction to the
conductivity. a) The results for sample 1 (circles) are obtained by determining the WL correction using Eq.(\ref{eqn:WL}) (in region I the results for sample S3 are also displayed by triangles). b) For sample S2 the WL contribution is suppressed by magnetic field.  Solid lines are fits to
Eq.(\ref{eqn:EEI}).}
\label{fig:four}
\end{figure}

The extracted EEI correction is shown for samples S1
and S2 in Fig.~\ref{fig:four}, where we also add the result for
sample S3 in region I. It is indeed logarithmic in temperature, Eq.(\ref{eqn:EEI}), with close values of $A$, $A$ = 0.5 - 0.8, for all three regions of the carrier density in the studied samples.

To interpret the obtained value of $A$, we note that the theory \cite{AltshulerAronov} distinguishes between the contributions from different quantum states of two interacting electrons, commonly referred to as `channels'. The coefficient $A$ takes the form $A=1+c\left(1-\ln(1+F_{0}^{\sigma})/F_{0}^{\sigma}\right)$, where $F_{0}^{\sigma}$ is the Fermi-liquid constant. While the first term in this relation represents the universal contribution of the `singlet' channel, the second (Hartree) term describes the contributions of $c$ `triplet' channels. For example, in a single-valley 2D system (such as in GaAs) the coefficient $c=3$ due to identical contributions of three spin triplet states. (When the this degeneracy is lifted by magnetic field \cite{AltshulerAronov}, two components become suppressed, resulting in $c=1$.)

In two-valley 2D systems (e.g., in Si-MOSFETs \cite{Vitkalov, Gershenson})
the situation is more complicated. In the absence of
inter-valley scattering, the valley index $v=\pm$ is a good quantum
number. In this case the overall number of channels is $16$, due to
four-fold spin degeneracy of two interacting electrons and an additional four-fold degeneracy due to the two valleys.
This gives the prefactor $c=15$. This result also holds if the inter-valley scattering is weak, $k_{B}T\gg\hbar/\tau_i$, i.e. when the typical electron energy is larger than the characteristic rate of inter-valley scattering. However, at low temperatures, $k_{B}T\ll\hbar/\tau_i$, strong inter-valley scattering mixes the valleys and $A$ takes the same form as in the single-valley case.

Unlike in Si-MOSFETs, in graphene the valley dynamics is governed by two
characteristic times, the inter-valley scattering time $\tau_i$ and intra-valley dephasing time $\tau_{*}$ \cite{McCannPRL06}.
In our experiments, the inter-valley scattering rate  $\hbar/\tau_i$ is of the order of 3 K, while the intra-valley dephasing rate is above 20 K (see Table \ref{tab:Parameters}). Thus the intermediate regime, $\hbar/\tau_{i}<k_{B}T<\hbar/\tau_{*}$, becomes possible. In this case, the channels with two electrons from different valleys give no contribution. (This situation is similar to the one which occurs for universal conductance fluctuations in graphene \cite{EfetovFalko}.) As there are two spin states per electron and two states for two electrons in the same valley, there are eight remaining channels, one of which is both spin and valley singlet, so that $c=7$. Thus we arrive at the following expression for $A$ in Eq.(\ref{eqn:EEI}):
\begin{eqnarray}
A(F_{0}^{\sigma})=1+7\left(1-\ln(1+F_{0}^{\sigma})
/F_{0}^{\sigma}\right)~.
\label{eqn:A}
\end{eqnarray}
We have confirmed this analysis by standard diagrammatic calculations, where we used a common assumption that all channels except for the singlet are described by the same Fermi-liquid parameter.

Using Eq.(\ref{eqn:A}) and the experimental values of $A$,
Fig. \ref{fig:four}, we find the values of $F_{0}^{\sigma}$ to be between
-0.08 and -0.13. It is interesting to note that the value of  $F_{0}^{\sigma}$ found in GaAs and Si systems at $r_s\sim1$ is between -0.15 and -0.2 \cite{Gershenson, Proskuryakov1, Vitkalov}. To explain this relatively low value of~$F_0^\sigma \approx -0.1$ found in our experiments, we note that in graphene this constant is suppressed due to the chirality of charge carriers, which prevents large-angle electron-electron scattering.
Indeed, in the non-chiral 2DEG, the constant~$F_0^\sigma$
can be found by averaging the electron-electron scattering
amplitude over all possible scattering angles (see, e.g.,
\cite{AltshulerAronov}):
$F_0^\sigma = -\nu \langle U(|{\bf p} - {\bf p}'|)\rangle$.
Here~$U(q)$ is the Fourier component of the interaction potential,
and $\nu$
is the density of electron states per spin/valley. In a chiral system, the scattering amplitude for each electron is suppressed by the factor~$\cos(\theta/2)$, where~$\theta$ is the
scattering angle (see, e.g. \cite{Cheianov}), so that
$F_0^\sigma = - \nu \langle U(|{\bf p} - {\bf p}'|) \cos^2(\theta/2) \rangle$.

For a simple estimate away from the Dirac point,
we use the Thomas-Fermi approximation for the
interaction potential, $U(q) = 2\pi e_\ast^2 / (q + \kappa)$,
with the effective charge, $e_\ast^2 = 2e^2 /(\epsilon + 1)$
that includes suppression of the Coulomb interaction by ${\rm SiO}_2$ substrate,
$\epsilon = 3.9$.
The screening
parameter~$\kappa = 4\cdot 2\pi \nu e_\ast^2$ includes contributions
from four degenerate single-electron states (in graphene the density of electron states is ~$\nu = \epsilon_F / 2\pi v_F^2$). This gives
\begin{equation}
F_0^\sigma = - \alpha \int\limits_{0}^{\pi}\frac{d\theta}{\pi}
\frac{\cos^2\frac{\theta}{2}}{2\sin\frac{\theta}{2} + 2 \alpha}\ ,
\label{eqn:F}
\end{equation}
where~$\alpha = e_\ast^2/\hbar v_F \approx 0.88$ is the
dimensionless interaction constant (it is related to the parameter $r_s$ used in \cite{ZNA, Proskuryakov1, Shashkin, Gershenson1, Vitkalov, Gershenson} as $r_s=\sqrt{2} \alpha$).
Evaluating this integral, we find~$F_0^\sigma = -0.10$, which is in agreement with our measurements.
Note, that a similar calculation for a non-chiral electron liquid with two valleys gives a larger value ~$F_0^\sigma \approx -0.19$
for the same value of~$\alpha$.  Approximation (\ref{eqn:F}) which neglects effects of strong interaction, such as the Fermi velocity and $Z$ factor renormalisations \cite{MacDonald}, is expected to be valid for $\alpha\leq1$, which is the case for graphene. Our result is in agreement with the value of $F_{0}^{a}$ in \cite{MacDonald} for the studied range of charge densities. (To compare $F_{0}^{\sigma}$ with $F_{0}^{a}$ in \cite{MacDonald}, one has to take into account that these quantities are related as $F_{0}^{a}=2F_{0}^{\sigma}$.)

In summary, we show that electron-electron interaction plays an important role in the low-temperature conductivity of carriers in graphene. Unexpectedly for the EEI correction, its magnitude is affected by the intra-valley decoherence rate due to elastic scattering. We find the value of the interaction parameter $F_{0}^{\sigma}$ in graphene, which is lower than in other 2D systems studied earlier.

We are grateful to V. I. Fal'ko, A. D. Mirlin, E. McCann, I. V. Lerner and M. Polini for useful
discussions and to R. V. Gorbachev and F. V. Tikhonenko for support in fabricating samples.


\begin{thebibliography}{99}

\bibitem{Bergman} G. Bergman, Phys. Rep. \textbf{107,} 1 (1984).
\bibitem{AltshulerAronov}B. L. Altshuler and A. G. Aronov, \textit{Electron-Electron Interactions
in Disordered Systems}, edited by A. L. Efros and M. Pollak (North-Holland, Amsterdam, 1985).
\bibitem{ZNA} G.~Zala, B.~N.~Narozhny, and I.~L.~Aleiner, Phys. Rev. B \textbf{64}, 214204 (2001).
\bibitem{Proskuryakov1} Y. Y. Proskuryakov \textit{et al.}, Phys. Rev. Lett. \textbf{89}, 076406 (2002);
A. K. Savchenko \textit{et al.}, Phys. Stat. Sol. (b) \textbf{242}, 1204 (2005).
\bibitem{Shashkin} A. A. Shashkin \textit{et al.}, Phys. Rev. B \textbf{66}, 073303 (2002)
\bibitem{Gershenson1} V. M. Pudalov \textit{et al.}, Phys. Rev. Lett. \textbf{91}, 126403 (2003).
\bibitem{Vitkalov} S. A. Vitkalov \textit{et al.}, Phys. Rev. B \textbf{67}, 113310 (2003).
\bibitem{Gershenson} N. N. Klimov \textit{et al.}, Phys. Rev. B \textbf{78}, 195308 (2008).
\bibitem{Ando} H. Suzuura and T. Ando, Phys. Rev. Lett. \textbf{89}, 266603 (2002).
\bibitem{McCannPRL06} E. McCann \textit{et al.}, Phys. Rev. Lett. \textbf{97}, 146805 (2006).
\bibitem{Tikhonenko} F.~V.~Tikhonenko \textit{et al.}, Phys. Rev. Lett. \textbf{100}, 056802 (2008).
\bibitem{TikhonenkoWAL} F. V. Tikhonenko \textit{et al.}, Phys. Rev. Lett. \textbf{103}, 226801 (2009).
\bibitem{Cheianov} V. V. Cheianov and V. I. Fal'ko, Phys. Rev. Lett. \textbf{97}, 226801 (2006).
\bibitem{Ando98} T. Ando, T. Nakanishi, and R. Saito, J. Phys. Soc. Jpn. {\bf 67}, 2857 (1998).
\bibitem{Monolayer} K. S. Novoselov \textit{et al.}, Nature \textbf{438}, 197 (2005).
\bibitem{Morozov} S. V. Morozov \textit{et al.}, Phys. Rev. Lett. \textbf{100}, 016602 (2008).
\bibitem{Fuhrer} J. H. Chen \textit{et al.}, Nature Nanotechnology \textbf{3,} 206 (2008).
\bibitem{Acoustic} T.~Stauber, N.~M.~R.~Peres, and F.~Guinea, Phys. Rev. B \textbf{76}, 205423 (2007); E.~H.~Hwang and S.~Das Sarma, Phys. Rev. B \textbf{77}, 115449 (2008).
\bibitem{Btr} H.-P. Wittmann and A. Schmid, J. Low Temp. Phys. \textbf{69}, 131 (1987); Y. Y. Proskuryakov \textit{et al.}, Phys. Rev. Lett. \textbf{86}, 4895 (2001); K. E. J. Goh, M. Y. Simmons, and A. R. Hamilton, Phys. Rev. B \textbf{77}, 235410 (2008).
\bibitem{Lande} Y. Zhang \textit{et al.}, Phys. Rev. Lett. \textbf{96}, 136806 (2006).
\bibitem{EfetovFalko} M. Yu. Kharitonov and K. B. Efetov, Phys. Rev. B \textbf{78}, 033404 (2008); K. Kechedzhi, O. Kashuba, and V. I. Fal'ko, Phys. Rev. B \textbf{77}, 193403 (2008).
\bibitem{MacDonald} M. Polini \textit{et al.}, Solid State Commun. \textbf{143}, 58 (2007).


\end{thebibliography}
\end{document}